\documentclass[]{interact}

\usepackage{hyperref}
\usepackage{epstopdf}
\usepackage[caption=false]{subfig}
\usepackage{xcolor}

\usepackage{natbib}
\bibpunct[, ]{(}{)}{;}{a}{}{,}
\renewcommand\bibfont{\fontsize{10}{12}\selectfont}

\theoremstyle{plain}

\theoremstyle{definition}

\theoremstyle{remark}

\begin{document}


\title{Multi-Label Classification of Thoracic Diseases using Dense Convolutional Network on Chest Radiographs}

\author{
\name{Dipkamal Bhusal\textsuperscript{a} and Sanjeeb Prasad Panday\textsuperscript{a}\thanks{CONTACT Sanjeeb Prasad Panday Email: sanjeeb@ioe.edu.np}}
\affil{\textsuperscript{a}Department of Electronics and Computer Engineering, Institute of Engineering, Pulchowk Campus, Lalitpur, Nepal
}
}

\maketitle

\begin{abstract}
Traditional methods of identifying pathologies in X-ray images rely heavily on skilled human interpretation and are often time-consuming. The advent of deep learning techniques has enabled the development of automated disease diagnosis systems. Still, the performance of such systems is opaque to end-users and limited to detecting a single pathology. In this paper, we propose a multi-label disease prediction model that allows the detection of more than one pathology at a given test time. We use a dense convolutional neural network (DenseNet) for disease diagnosis. Our proposed model achieved the highest AUC score of 0.896 for the condition Cardiomegaly with an accuracy of 0.826, while the lowest AUC score was obtained for Nodule, at 0.655 with an accuracy of 0.66. To build trust in decision-making, we generated heatmaps on X-rays to visualize the regions where the model paid attention to make certain predictions. Our proposed automated disease prediction model obtained highly confident high-performance metrics in multi-label disease prediction tasks. We believe this work will contribute towards the development of reliable and trustworthy automated diagnosis systems for disease.
\end{abstract}


\section{Introduction}\label{sec:intro}

Deep learning has achieved remarkable performance in various image classification tasks largely due to the availability of labeled datasets (\cite{NIPS2012_c399862d, ren2015faster, simonyan2014very, szegedy2017inception, he2016deep}). Deep learning has also shown immense potential in health analytics, particularly in automating the diagnosis process. This is especially important given that approximately 3 billion people lack access to medical imaging expertise, as reported by the World Health Organization (\cite{survey}). An automated diagnosis system can be particularly beneficial in areas where medical expertise is limited.

\begin{figure}[ht]
\begin{minipage}{0.2\linewidth}
\includegraphics[width=\textwidth]{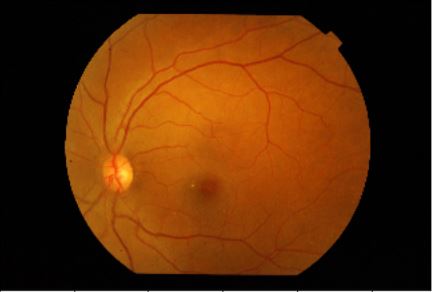}
\caption [A sample of fundus photo] { A sample of fundus photo  (\cite{eye})} 
\label{fig:figure1}
\end{minipage}%
\hfill
\begin{minipage}{0.2\linewidth}
\includegraphics[width=\textwidth]{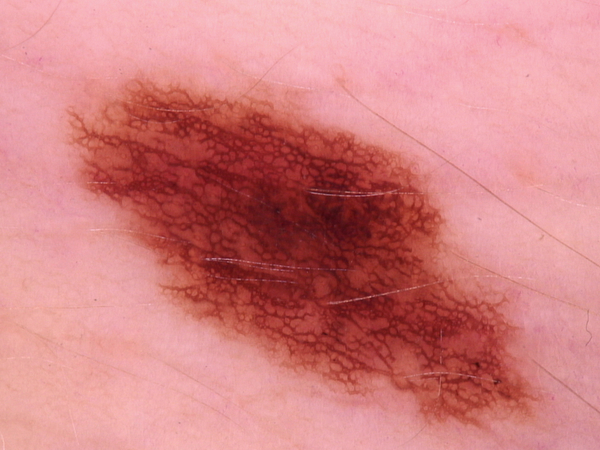}
\caption [A sample of skin image] {A sample of skin image (\cite{skin})}
\label{fig:figure2}
\end{minipage}%
\hfill
\begin{minipage}{0.2\linewidth}
\includegraphics[width=\textwidth]{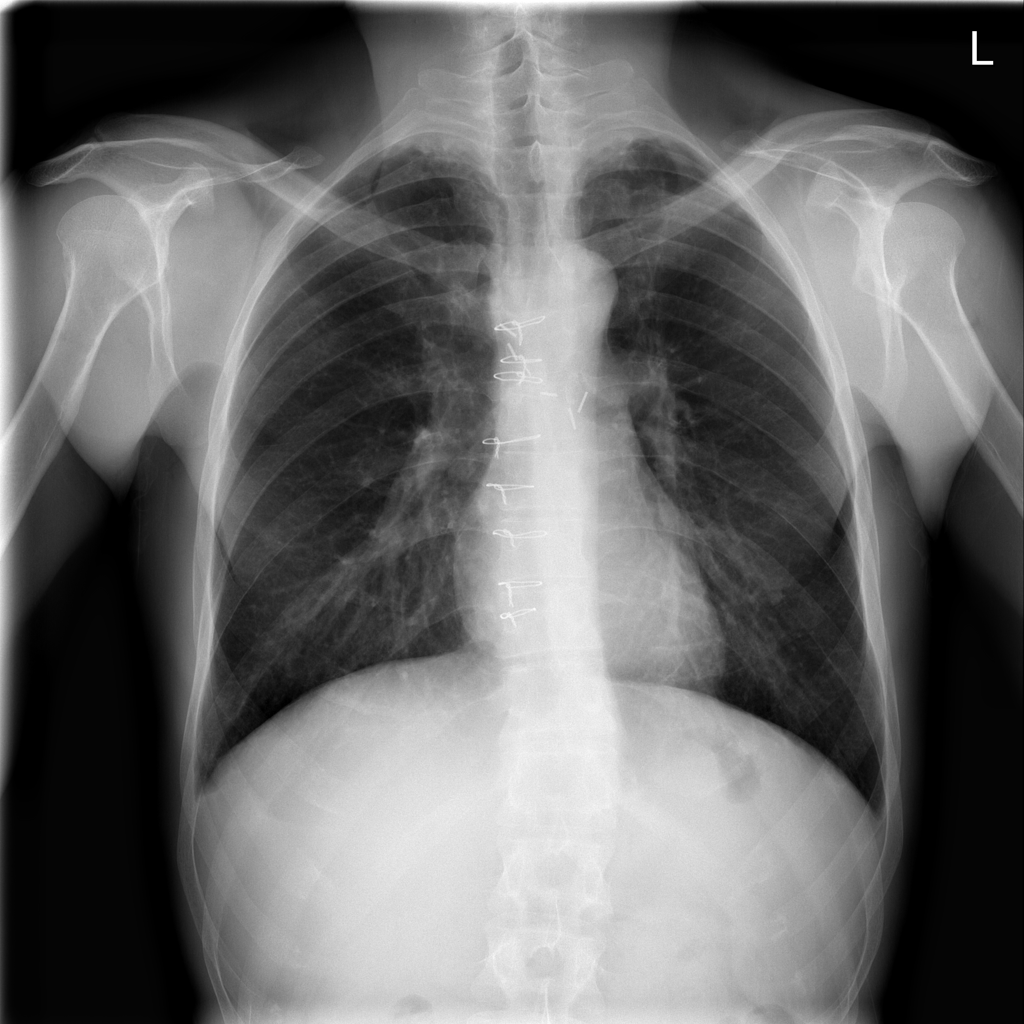}
\caption [A sample of chest x-ray ] {A sample of chest x-ray (\cite{ChestX-Ray8})}
\label{fig:figure3}
\end{minipage}
\end{figure}

Given a medical image of a patient as input, a disease prediction system provides the probability of the occurrence of a disease. This approach represents a single-label classification problem. Examples of such diagnoses include diabetic retinopathy in eye fundus images, skin cancer in skin lesion images, and pneumonia in chest X-rays (Figure \ref{fig:figure1}, Figure \ref{fig:figure2}, and Figure \ref{fig:figure3}). However, in certain cases, multi-label prediction becomes crucial as it provides the probabilities of multiple pathologies occurring within the same medical image. This is particularly important when there are possibilities of more than one disease being present.

Thoracic diseases pose a significant threat to the global population, with recent events such as the COVID-19 pandemic causing respiratory illnesses worldwide (\cite{he2020coronavirus}). In addition to COVID-19, chest radiography plays a crucial role in screening common thoracic diseases, including pneumonia, cardiomegaly, and pneumothorax. It is estimated that more than 2 billion chest radiography procedures are performed annually (\cite{raoof2012interpretation}). However, the increasing workload for radiologists has led to inefficiencies in disease identification due to fatigue, potentially resulting in cognitive or perceptual errors in diagnosis. Consequently, there is a growing interest in first-world countries to develop computer-aided diagnosis systems that can assist medical professionals in evaluating X-rays. These systems have the potential to assist medical professionals and reduce diagnostic errors.

\par 

Most existing studies on disease diagnosis using chest X-rays primarily focus on detecting a single pathology, such as pneumonia or COVID-19 (\cite{2015chest, 2017chest, chexnet, dasanayaka2021deep, hussain2023performance}). However, an X-ray image can exhibit multiple pathological conditions simultaneously. Detecting multiple pathologies can provide a comprehensive view of the patient's health from a single image. Single-label classifications may produce false negatives when patients have multiple diseases, as they focus solely on the primary condition. Multi-label classification can help reduce false negatives by identifying secondary or co-occurring diseases. Multi-label classification can also be valuable in epidemiological studies and public health research. It can provide insights into the prevalence and co-occurrence of diseases in specific populations, aiding in resource allocation and healthcare planning. In this research, we employ a 121-layer DenseNet architecture to perform diagnostic predictions for 14 distinct pathological conditions in chest X-rays. Additionally, we utilize the GRADCAM explanation method to localize specific areas within the chest radiograph to visualize the regions to which the model paid attention to make disease predictions, enhancing our understanding of the model's predictions. The detection of these 14 different pathology conditions, including `Atelectasis', `Cardiomegaly', `Consolidation', `Edema', `Emphysema', `Effusion', `Fibrosis', `Hernia', `Infiltration', `Mass', `Nodule', `Pneumothorax', `Pleural Thickening', and `Pneumonia', presents a multi-label classification problem. The input to the DenseNet architecture is a chest X-ray image; the output is a label that provides the probability of each pathology being present in the X-ray. The code for our approach is available on Github\footnote{\url{https://github.com/dipkamal/chestxrayclassifier}}.



\textit{Our paper is structured as follows:} We begin with providing the necessary background information in Section \ref{sec:background}. This section covers a comprehensive overview of deep learning and highlights the existing works in the field of disease prediction and model interpretability.

Section \ref{sec:method} explains our methodology, including the dataset used, the architecture of our model, and the various steps involved in developing the model. We also introduce the evaluation criteria utilized in assessing the performance of our proposed diagnostic model.

In Section \ref{sec:results}, we present the evaluation results of our diagnostic model. We analyze and interpret the findings and present various quantitative results of disease prediction, showcasing the accuracy of our model in detecting different pathologies. We also display the qualitative results of model interpretation, demonstrating the effectiveness of our model in generating meaningful visualizations.

Finally, we conclude our paper with limitations in Section \ref{sec:limitations} and conclusions in Section \ref{sec:conclusion}, summarizing our key findings and contributions. We also discuss the implications of our work and highlight future directions for research in the field of thoracic disease prediction using deep learning techniques.

\section{Background}\label{sec:background}
\subsection{Convolutional Neural Network}
Convolutional Neural Networks (CNNs) \cite{fukushima1991handwritten} are widely used neural network architectures in image classification tasks. They efficiently extract and learn image features through convolution and pooling layers. The pioneering CNN architecture, AlexNet \cite{NIPS2012_c399862d}, employed multiple convolutional and fully connected layers, achieving state-of-the-art performance on the ImageNet dataset. VGG Net \cite{simonyan2014very} further improved upon AlexNet by introducing deeper models with 16 or 19 weight layers, known as VGG16 and VGG19, respectively. However, increasing the depth of CNNs can lead to overfitting \cite{goodfellow2016deep}. To address this, Inception Net \cite{szegedy2017inception} proposed using filters of different sizes within the same level to widen the network rather than making it deeper. Residual Networks (ResNets) \cite{he2016deep} introduced skip connections to enable the training of even deeper models. DenseNet \cite{densenet} parallelized this approach by connecting each layer to all preceding and succeeding layers, addressing the vanishing gradient problem in deep neural networks.

Although these architectures differ in their topology for transmitting features across layers, they share the fundamental CNN principle of convolution, sub-sampling, dense, and softmax layers. In the convolution layer, filters are applied to the input image to extract features. The sub-sampling layer reduces the spatial size of the output from the convolution layer. The dense layer is a fully connected layer that processes the output from the sub-sampling layer. Finally, the softmax layer computes the probability distribution over the output classes.

\subsection{Deep learning for thoracic disease predictions}
Deep learning models have shown notable improvements in diagnosis accuracy across various medical imaging applications, including detecting diabetic retinopathy \cite{eye} and skin cancer \cite{skin}. Given the substantial impact of thoracic diseases on public health and the widespread use of chest X-rays in medical diagnosis, numerous research projects have explored the performance of deep learning in detecting these conditions.

In 2015, \cite{2015chest} used a pre-trained image classifier for classifying pathologies in chest radiographs, demonstrating the feasibility of detecting X-ray pathology\cite{donahue2014decaf}. In 2017, \cite{2017chest} presented a similar CNN classifier that achieved an AUC of 0.964 using a medium-sized dataset of 35,000 X-rays annotated by 2443 radiologists. The authors achieved an overall sensitivity and specificity of 91\% using GoogleNet \cite{szegedy2015going}. \cite{TB1} evaluated the performance of CNNs in tuberculosis detection using a small dataset of 1007 chest X-rays. They experimented with pretrained and untrained versions of two architectures, AlexNet \cite{NIPS2012_c399862d} and GoogleNet \cite{szegedy2015going}, and obtained the best performance with an ensemble of both architectures in the pretrained condition (AUC = 0.99). The pretrained models consistently outperformed the untrained models. Similarly, \cite{TB2} compared the performance of a computer-aided tuberculosis diagnosis system (CAD4TB) with that of health professionals and found that the tuberculosis assessment of CAD4TB was comparable to that of health officers. In 2016, \cite{ChestX-Ray8} proposed weakly controlled multi-label classification and localization of thoracic diseases using deep learning. 
In 2017, \cite{chexnet} designed a deep learning model called CheXNet, which utilized a 121-layer CNN with dense connections and batch normalization to detect pneumonia. The model was trained on a publicly available dataset of 100,000 chest X-ray images and outperformed the average radiologist performance. 
\cite{bar2018chest} used a pretrained model on a non-medical dataset and fine-tuned it on pathology features for disease identification. \cite{dasanayaka2021deep} presented deep learning-based segmentation techniques to detect pulmonary tuberculosis. \cite{patel2023machine} utilized the Littlewood-Paley Empirical Wavelet Transform (LPEWT) to decompose lung images into sub-bands and extract robust features for lung disease detection. 
Deep learning has also been extensively applied in the detection of COVID-19\cite{bhuyan2022covid, farooq2020covid, yang2020chest, li2020ct, pushparaj2022detailed, irene2022efficient, dhruv2023hybrid}. 


\textbf{Limitations:} Most disease prediction models focus on single-label classification, where the model only detects the presence of a single pathology. However, multi-label disease classification can offer several advantages over single-label classification. Multi-label diagnosis is akin to realistic representation since, in clinical practice, it's common for patients to have multiple medical conditions. Multi-label classification allows a single instance (e.g., an x-ray image) to be associated with multiple disease labels. This provides a more comprehensive view of the patient's health, as many patients may suffer from multiple medical conditions simultaneously. Single-label classification may force a medical professional to decide which disease is the ``primary" one when a patient has multiple conditions. This can lead to information loss, as secondary conditions may be overlooked. A multi-label classification doesn't require this decision and captures all relevant conditions. 

\subsection{Deep learning interpretability}


Deep learning models are examples of black box predictors that often lack interpretability. End-users struggle to understand the complex inner workings of deep neural networks and the reasoning behind their output predictions. However, understanding the underlying process of a model's prediction is crucial for building trust with users \cite{ribeiro2016should}. When users need to make decisions based on a model's predictions, having insights into the reasons behind those predictions becomes essential. Therefore, explanations from deep learning models are vital for establishing trust and confidence in their outputs \cite{pieters2011explanation}.

Model designers can also utilize explanations derived from black box models to verify if the model functions as intended. Moreover, model users can employ explanation methods to feel more comfortable and confident when using black box models by gaining information about the model's predictions on specific test samples \cite{Bhusal2022SoKME}. Interpreting black box models also plays a significant role in evaluating factors such as fairness, privacy, reliability, causality, and trust \cite{doshi2017towards}. Consequently, developing explainable AI models is essential for ensuring transparency and accountability in decision-making processes.

Post-hoc explanation is a well-studied approach that focuses on explaining individual predictions made by a model rather than providing a comprehensive understanding of the entire decision-making process \cite{bodria2021benchmarking}. In the context of a black box model $F(x)$, where $x = {x_1, x_2, ..., x_N}$ represents the input and $F(x) = y$ is the model's prediction, a post hoc explanation strategy denoted as $\gamma(x)$ is employed to generate an explanation vector $E_k(x)$ that indicates the relevance or importance of the $k$ given features. 

\begin{figure}
\centering
\resizebox*{10cm}{!}{\includegraphics{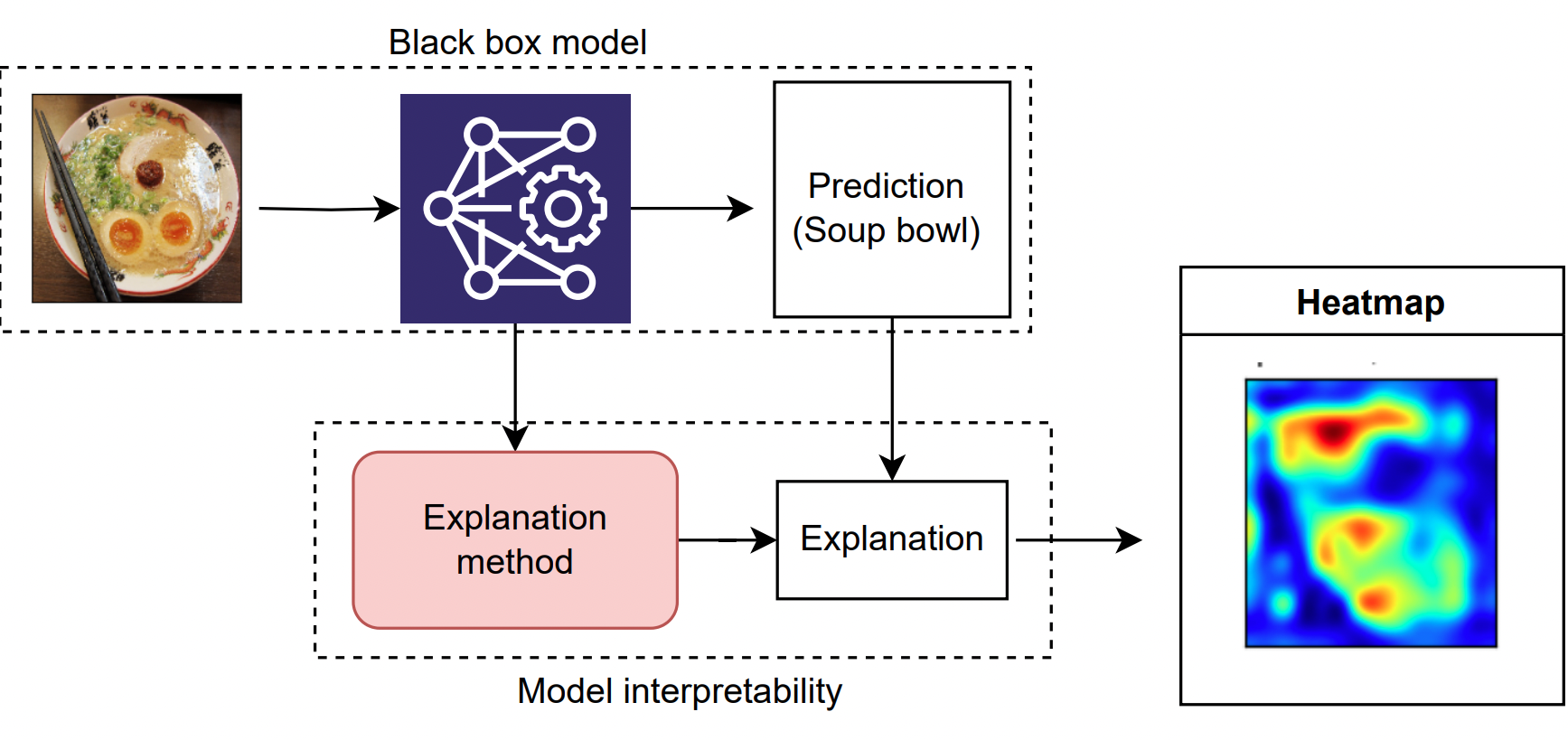}}
\caption{Pixel attributions or saliency maps for an image-classifier test case using Grad-CAM \cite{molnar2022}} \label{fig:interpret}
\end{figure}

In the case of image data, pixel attribution methods are commonly used for post-hoc explanations. These methods aim to highlight the pixels in an image that significantly contribute to a specific classification made by the model. By visualizing these highlighted pixels, it becomes possible to understand which regions of the image played a crucial role in the model's prediction. Figure \ref{fig:interpret} provides an overview of post-hoc explanations in image classifiers. We provide a brief overview of some major post-hoc explanation methods below:

\begin{enumerate}
    \item \textbf{Perturbation-based methods:} Perturbation-based methods involve perturbing specific regions of the input and observing the impact on the model's output to assess the importance of different regions. Two popular perturbation-based post-hoc explanation methods are Local Interpretable Model-Agnostic Explanations (LIME) \cite{ribeiro2016should} and Shapley Additive Explanation (SHAP) \cite{lundberg2017unified}. LIME creates an interpretable surrogate model by perturbing a test instance and generating new data samples, using linear regression to explain the predictions. The weights of the surrogate model indicate the importance of features. SHAP uses a game-theoretic approach to compute Shapley values for each feature, generating new samples around a given instance and fitting an interpretable linear model. However, unlike LIME, SHAP weights the new instances according to the weight a coalition would receive in the Shapley value estimation rather than their closeness to the original sample. SHAP provides reliable and consistent explanations with a theoretical foundation, while LIME is faster and easier to implement. Both methods successfully provide insights into black box models, allowing users to understand specific predictions. However, they are primarily suited for tabular data and not commonly used for unstructured data like images \cite{slack2020fooling}.
    
    \item \textbf{Gradient-based methods:} Gradient-based methods calculate the gradients of the model's output with respect to the input features to measure their importance. The Gradient method \cite{simonyan2013deep} computes the gradient of the model output with respect to the input features, providing pixel attributions. Integrated Gradient (IG) \cite{sundararajan2017axiomatic} accumulates gradients along a linear path from a baseline to the test sample, addressing some limitations of the Gradient method. SmoothGrad \cite{smilkov2017smoothgrad} and NoiseGrad \cite{bykov2022noisegrad} are improvements over gradient-based methods. SmoothGrad adds Gaussian noise to generate multiple samples, averaging their pixel attributions. NoiseGrad introduces noise to model parameters, generating multiple models and averaging feature attributions. GradientSHAP \cite{erion2021improving} combines SHAP and SmoothGrad with integrated gradients, selecting a baseline randomly and averaging the resulting attributions. Grad-CAM \cite{gradcam} and Grad-CAM++ \cite{8354201} are class activation map methods that compute feature-importance maps of convolutional layers with respect to specific classes. These methods backpropagate gradients from the model prediction to the convolutional layers, obtaining coarse localization maps of feature attribution. Gradient-based methods are popular due to their simplicity, speed, and effectiveness in providing feature attributions, particularly in image classification tasks.
\end{enumerate}

\section{Methodology}\label{sec:method}

\subsection{Data preparation}

\subsubsection{Addressing Data Leakage}\label{sec:dataleakage}
To prevent data leakage in medical image analysis, we took precautions to ensure that images from the same patient were not present in both the training and test sets. Data leakage can occur when the model sees images of the same patient during training and testing, leading to biased results \cite{rathore2017review}.

We utilized the ChestX-ray8 dataset \cite{ChestX-Ray8} as our primary dataset and randomly selected 99,000 images. Random selection of datasets introduces an element of stochasticity that helps a model learn from different parts of data distribution for robust representation. However, we implemented a patient-level dataset division to obtain the train and test set to avoid data leakage. This means that all images of a particular patient were assigned exclusively to either the training or test set, but not both. By separating the images at the patient level, we minimized the data leakage risk and avoided introducing biases into our deep learning model.

This approach helps maintain the integrity of the evaluation process, ensuring that the model generalizes well to unseen patients and accurately reflects its performance in real-world scenarios.

\subsubsection{Addressing Class Imbalance}

\begin{figure}
\centering
\subfloat[Highly skewed data distribution before adjusting class imbalance]{%
\resizebox*{6cm}{!}{\includegraphics{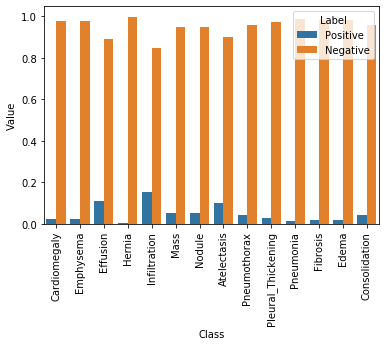}}}\hspace{5pt}
\subfloat[After adjusting class imbalance problem in the dataset]{%
\resizebox*{6cm}{!}{\includegraphics{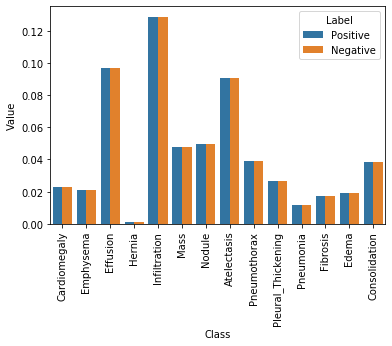}}}
\caption{Solving class-imbalance problem} \label{fig:class-imb-problem}
\end{figure}

The class imbalance problem in the dataset refers to the significant difference in the number of positive cases (images with diseases) compared to negative cases (images without diseases). This imbalance can pose challenges during model training, as the algorithm may prioritize the majority class and overlook the minority class, leading to biased predictions.

Figure \ref{fig:class-imb-problem} illustrates the class imbalance problem in our dataset, highlighting the low proportion of positive cases compared to negative cases for certain pathological conditions. For example, the Hernia class has a ratio of positive to negative cases of approximately 0.02, indicating a highly imbalanced distribution. Similarly, the Infiltration class, which has the highest number of positive labels, exhibits a ratio of around 0.18.

Various techniques can be employed to mitigate class imbalance. {One common approach is to assign different weights to the positive and negative classes during the loss calculation. By giving more importance to the minority class (positive cases) in the loss calculation, the model is encouraged to focus on correctly predicting these cases. We employ this technique in our study and modify the cross-entropy loss function.} Other methods for addressing class imbalance include oversampling the minority class, undersampling the majority class, or using a combination of both. These techniques create a balanced training set by either replicating instances of the minority class or reducing instances of the majority class.

The normal cross-entropy loss, which is commonly used in classification models, for the $i^{th}$ example is given by:

\begin{equation}
L_{cross-entropy}(x_i)=-(y_i \log(g(x_i)) + (1-y_i) \log(1-g(x_i)))
\label{eq:cross}
\end{equation}

Here, $x_i$ and $y_i$ denote the features and label of the given image, respectively, and $g(x_i)$ represents the model prediction. Either $y_i$ or $(1-y_i)$ will contribute to the loss at any given time since when $y_i$ equals one, $(1-y_i)$ is zero and vice versa. This means that one label will dominate the loss in an imbalanced dataset. For an entire training set of size $N$, the cross-entropy loss is given by:

\begin{equation}
\begin{split}
L_{cross-entropy}(D)=-(1/N) ( \sum_{\text{positive examples}} \log (g(x_i)) +\ \sum_{\text{negative examples}} \log(1-g(x_i)))
\end{split}
\label{eq:balcross}
\end{equation}

Here, the first summation term represents the loss for all positive examples, while the second summation term represents the loss for all negative examples. This loss function leads to a bias towards the majority class in an imbalanced dataset.

To address the problem of highly skewed data distribution, ensuring that each class's labels make an equal contribution is crucial. This can be achieved by multiplying each example from each class by a class-specific weight factor, denoted as $w_{pos}$ and $w_{neg}$. To obtain equal contribution from both positive and negative classes, we aim to satisfy the following conditions:

\begin{equation}
w_{pos} \times freq_{p} = w_{neg} \times freq_{n}
\end{equation}

For this condition to hold, we compute weight factors that are determined based on the frequency of positive and negative examples in the dataset:

\begin{equation}
w_{pos} = freq_{neg}
\end{equation}

and

\begin{equation}
w_{neg} = freq_{pos}
\end{equation}

Here, $freq_{p}$ and $freq_{n}$ represent the frequency of positive and negative examples, respectively, defined as:

\begin{equation}
freq_p=\text{(number of positive examples)}/N
\end{equation}

and

\begin{equation}
freq_n=\text{(number of negative examples)}/N
\end{equation}

By adjusting the class imbalance, we achieve a balanced contribution of positive and negative labels to the loss function as shown in Figure \ref{fig:class-imb-problem}. Therefore, the final weighted loss after computing the positive and negative weights is given by:

\begin{equation}\label{eqn:los}
\begin{split}
{L}{cross-entropy}^{w}(x) = - (w_{pos} y \log(f(x)) + w_{neg}(1-y) \ \log( 1 - f(x) ) )
\end{split}
\end{equation}

Here, $y$ represents the true label, $f(x)$ represents the predicted label, and $w_{pos}$ and $w_{neg}$ represent the class-specific weight factors for positive and negative examples, respectively.

\subsubsection{Pre-processing}
Several preprocessing steps were performed on the training dataset to prepare the images for training the deep convolutional network. These steps aimed to standardize the data distribution and make it compatible with the chosen architecture and pre-trained model \cite{abdou2022literature}.

The first step in preprocessing involved normalization of the mean and standard deviation of the input data. This normalization process ensures that the pixel values across the images have a standardized distribution, which can help improve the training process and model convergence.

{Next, the x-ray images in the dataset have different sizes and were resized to a uniform dimension of 320 by 320 pixels.} This size is a suitable dimension for a deep convolutional network architecture used in the study. Resizing the images to a specific dimension ensures consistency in the input size for the model.

To facilitate transfer learning, we utilized a pre-trained model from ImageNet. However, the pre-trained model was trained on RGB images, while the Chest X-ray images in the dataset are single-channel grayscale images. To overcome this, we converted the 1-channel X-ray images to a 3-channel format. This conversion involved duplicating the grayscale image to create three identical channels, mimicking the RGB format required by the pre-trained model. This step enables the utilization of the pre-trained model's learned features and weights.

In addition to preprocessing the training data, the test data also underwent normalization. The normalization process used the statistics (mean and standard deviation) calculated from the training set. Normalizing the test data using the training set's statistics made the overall distribution of data during training and testing consistent. This ensures that the model's performance on the test set reflects its ability to generalize to new, unseen data.


\subsection{Network Architecture}\label{sec:network}

Neural networks consist of multiple layers that process input data to produce output predictions. Each layer takes a previous vector of activations, represented as $h(n)$, and applies a function $F$ to generate a new state activation, $h(n+1) = F(h(n))$, till the classifier layer is reached. However, simply adding more layers to a network does not always lead to improved performance. One of the major problems in deep neural networks is the vanishing gradient problem, which occurs when the gradients of the loss function decrease rapidly as they propagate backward. Eventually, these gradients become very small, preventing the weights from updating and stagnating the network.

The DenseNet architecture \cite{densenet} addresses the vanishing gradient problem by incorporating dense connections between layers. In DenseNet, each layer is directly connected to all the preceding layers in the network. This connectivity pattern enables each layer to access the feature maps produced by all the earlier layers. By facilitating the efficient flow of gradient signals and information throughout the network, DenseNet promotes more effective learning, improving accuracy. This connectivity scheme mitigates the vanishing gradient problem and promotes better information flow within the network, enabling deeper networks to be trained successfully. As a result, DenseNet has been shown to achieve state-of-the-art performance in various tasks, such as image classification and object detection.

\begin{figure}[t!]
\centering
\includegraphics[width=0.95\textwidth]{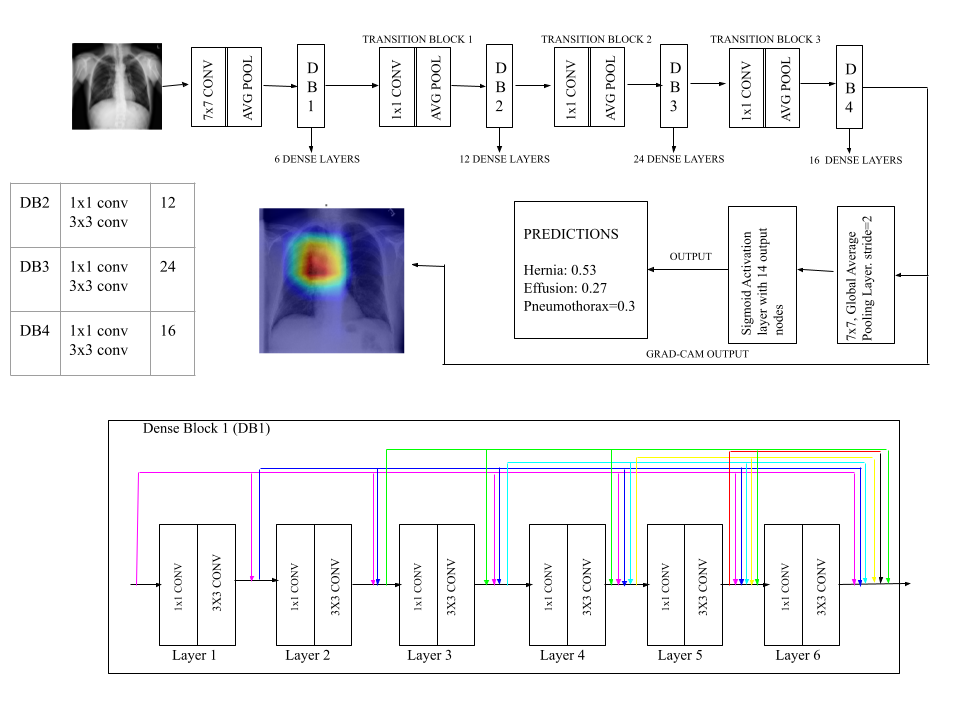}
\caption{Network architecture for the proposed diagnostic model using DenseNet.}
\end{figure}

In our project, we employed the DenseNet architecture, which comprises two main blocks: the DenseBlock and the TransitionBlock. The DenseBlock keeps the feature size dimension constant while varying the number of filters. Within a DenseBlock, each layer performs a 1x1 convolution for feature extraction and a 3x3 convolution to reduce the feature depth. On the other hand, the TransitionBlock is positioned between DenseBlocks and serves to downsample the feature size by applying a 1x1 convolution and a 2x2 average pooling function. This downsampling operation reduces spatial resolution, which helps control computational cost and prevent overfitting. The output of the TransitionBlock feeds into the next DenseBlock, repeating the process. This unique design of DenseNet enables efficient feature reuse and propagation, leading to improved accuracy and reduced overfitting. For our specific implementation, we used the DenseNet-121 architecture, which consists of four DenseBlocks. Dense Block 1 has six dense layers, Dense Block 2 has twelve dense layers, Dense Block 3 has twenty-four dense layers, and Dense Block 4 has sixteen dense layers. There are three TransitionBlocks between the DenseBlocks. The name ``DenseNet-121" indicates the total number of layers with trainable weights, which in this case is 121.

\subsection{Training}
To initialize the network, we employed transfer learning by utilizing pre-trained weights from ImageNet. The early layers of the DenseNet, which capture general image features like edges, were left unchanged. We skipped the top layers, which contain more specific image features, and added two additional layers: a Global Average Pooling layer and a Dense layer with sigmoid activation. The Global Average Pooling layer computes the average of the last convolutional layer's output, while the Dense layer with sigmoid activation provides predictions for all target classes. 
The loss function used in this project is the weighted loss function described by Equation \ref{eqn:los}. Once the neural network architecture was defined, we trained the model using back-propagation with mini-batch stochastic gradient descent. We used mini-batches of 32 images and the Adam optimizer with a default learning rate of 0.001. 


\section{Experiment and Results}\label{sec:results}

\subsection{Data}
We utilized the ChestX-ray8 dataset which consists of Chest X-ray images and is commonly used for research in the thoracic disease field. From this dataset, we randomly selected 99,000 images as our training set. Each image in the dataset was annotated with labels that identify 14 distinct pathological conditions. These labels provide information about the presence or absence of specific conditions in the X-ray images. Table \ref{tab:data} in the study provides a snapshot of the dataset, showing a summary of the dataset's characteristics, such as the pathological condition and patient ID. 

We addressed the data leakage as explained in Section \ref{sec:dataleakage} and created a train and test set. We employed deep learning architecture as explained in Section \ref{sec:network} to develop a model that can predict the presence or absence of the 14 pathological conditions based on the input Chest X-ray images. The training process involved feeding the model with the labeled images and adjusting its parameters to learn the patterns and features associated with each condition.

To evaluate the model's performance and assess its generalization capabilities, we created a separate test set. This test set comprised 500 images randomly selected from the test dataset. We applied the trained model to the test set and measured various evaluation criteria as explained in Section \ref{sec:criteria}.

\begin{table}[!t]
\tbl{Detail annotation in the dataset\textsuperscript{a}}
{\begin{tabular}{c c c c c c c c} \toprule 
Image & Atelectasis	& Consolidation	& Edema	& Effusion &	PatientId \\ \hline
015.png &	0 &	0 &	0 &	0  &	8270  \\ \hline 
001.png &	1  &	0 &	0  &	0 &	29855  \\ \hline
000.png &	0  &	0 &	0  &	0 &	1297  \\ \hline
002.png &	0 &		0 &	0  &	0 &	12359  \\ \hline
001.png &	0 &		0 &	0  &	0 &	17951  \\ \bottomrule
\end{tabular}}
\tabnote{\textsuperscript{a} All 14 pathological conditions are not displayed due to width constraint.}
\label{tab:data}
\end{table}

\subsection{Evaluation Metrics}\label{sec:criteria}
We computed several metrics to assess the generalization of our diagnostic models. These metrics include sensitivity, specificity, positive predictive value (PPV), negative predictive value (NPV), the receiver operating characteristic (ROC) curve, and the F1 score.

\textbf{Sensitivity}, also known as the \textbf{true positive rate} or \textbf{recall}, measures the probability that the model correctly predicts the presence of a disease given that the patient actually has the disease. It is computed as TP/(TP+FN), where TP represents the number of correctly predicted positive samples and FN represents the number of incorrectly predicted negative samples.

\textbf{Specificity}, also known as the \textbf{true negative ratio}, measures the probability that the model correctly predicts a patient as disease-free given that the patient is actually normal. It is computed as TN/(TN+FP), where TN represents the number of correctly predicted negative samples and FP represents the number of incorrectly predicted positive samples.

Diagnostically, sensitivity and specificity are not helpful alone. While sensitivity tells the probability that the test results positive given that the person already has the condition, the information of probability that the person has the disease given that the test gives positive is important. \textbf{Positive predictive value (PPV)}, also called \textbf{precision}, provides the probability that a patient has the disease, given that the model predicts a positive result. It is computed as TP/(TP+FP).

\textbf{Negative predictive value (NPV)} provides the probability that a patient does not have the disease when the model predicts a negative result. It is computed as TN/(TN+FN).

The \textbf{ROC curve }is a graphical representation of the true positive rate (sensitivity) versus the false positive rate (1-specificity) at different threshold settings for classification. The area under the ROC curve (AUC) is a threshold-independent measure of the model's goodness of fit and its ability to distinguish between positive and negative samples. In medical literature, this number also gives the probability that a randomly selected patient who experienced a condition had a higher risk score than a patient who had not experienced the event. A higher AUC indicates better performance of the model in distinguishing between positive and negative samples.

The \textbf{F1 score} is computed by taking the harmonic mean of precision and recall. It is a single metric that balances both precision and recall. The best possible value of the F1 score is 1 (perfect precision and recall), while the worst value is 0. If a single F1 score is required for multiclass classification, a micro-average (weighted by class frequency) or macro-average (same weights for all classes) approach can be used.

By analyzing these metrics, we gain insights into the model's performance in terms of sensitivity, specificity, predictive values, discrimination power (ROC curve), and overall classification accuracy (F1 score). These evaluations help us understand the strengths and limitations of the model in accurately predicting different pathologies.

\subsection{Evaluation of diagnostic model}

{We trained the model on different sizes of the training dataset with different training epochs to observe the generalization accuracy of the model against an unseen test set. Specifically, we wanted to analyze the following from our experiments: }

\begin{enumerate}
    \item {\textit{Effect of Dataset Size}: Data quantity can significantly impact a data-hungry deep learning model. By training the model with different dataset sizes, we can understand the relationship between data quantity and model performance. Typically, larger datasets lead to better generalization. However, datasets can over or underfit the model. Smaller datasets with smaller training epochs can underfit, whereas smaller datasets with higher training epochs are more prone to overfitting.}
    
    \item {\textit{Impact of Training Epochs}: The number of training epochs affects the convergence of a model by affecting how well the model converges to an optimal solution. Too few epochs may result in underfitting, while too many epochs may lead to overfitting.}
\end{enumerate}

\begin{figure}[t!]
\centering
\includegraphics[width=0.6\textwidth]{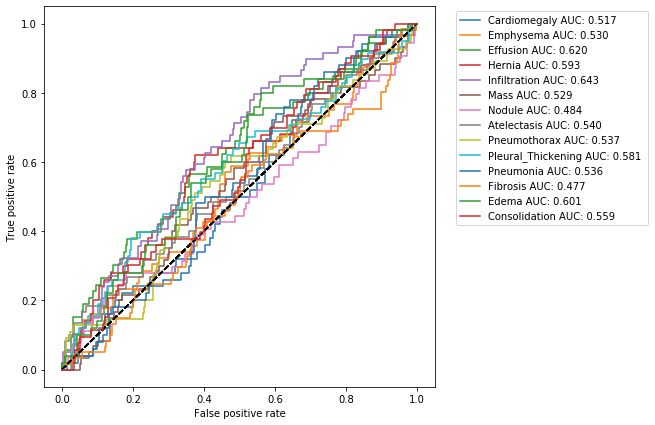}
\caption{ROC curve on the classification of thoracic pathologies by a model trained on a small dataset of 1000 images.}
\label{fig:ROC1000}
\end{figure}

The network was initially trained on a random dataset of 1000 images from the training set for five epochs. The model's performance was then assessed by plotting the receiver operating characteristic (ROC) curve, as shown in Figure \ref{fig:ROC1000}. The ROC curve illustrates the model's ability to distinguish between true positive and false positive rates for different classification thresholds.

From the results obtained, it can be observed that the model performed poorly, as the ROC curve for each pathology closely approximated the diagonal line. This suggests that the model acted as a poor classifier, exhibiting limited discriminative ability in predicting diseases accurately. Furthermore, the corresponding area under the ROC curve (AUC) scores further support this observation, indicating inaccurate disease prediction across various thresholds.

Based on these findings, we hypothesize that the underperformance of the model can be attributed to underfitting and the limited size of the training dataset, as well as the relatively small number of training epochs. With a small dataset, the model may not have had enough examples to learn the complex patterns and variations present in the medical images, leading to inadequate generalization to unseen data.

To address this issue and improve the model's performance, it is anticipated that training on a larger dataset and increasing the number of training epochs will yield better results. This would allow the model to capture a more comprehensive range of patterns and variations, leading to improved accuracy in disease prediction.

Table \ref{Tab:1000} shows the evaluation metrics for the model trained on 1000 X-ray images and tested on a test dataset. The results indicate poor generalization of the model, as reflected in its performance across all metrics. Although accuracy can be deceptive, particularly for conditions such as Pneumothorax, Hernia, and Pleural Thickening, the F1 scores provide a more reliable estimate of the model's ability to generalize, which is deemed unacceptable. Additionally, some conditions lack precision and recall values due to the absence of true positives and false positives. This highlights the limitations of the model and the need for further improvement.

\begin{table*} [h!]
    \caption{ Table of evaluation metrics after training on 1000 images}
    \resizebox{\textwidth}{!}{\begin{tabular}{l c c c c c c c c c}
    \hline
 & 	\textbf{Accuracy} &	\textbf{Prevalence} & \textbf{Sensitivity}	& \textbf{Specificity} &	\textbf{PPV}	 & \textbf{NPV} & \textbf{AUC} &	\textbf{F1} &	\textbf{Threshold}
\\ \hline

Cardiomegaly &	0.445 &	0.119 &	0.54 &	0.432 &	0.114 &	0.874 &	0.517 &	0.188 &	0.5 \\ \hline
Emphysema  &	0.864 &	0.133 &	0 &	0.997 &	0 &	0.866 &	0.53 &	0 &	0.5 \\ \hline
Effusion &	0.524 &	0.126 &	0.66 &	0.504 &	0.161 & 	0.911 &	0.62 &	0.259 &	0.5 \\ \hline
Hernia & 0.881 &	0.119 &	0 &	1 &	NaN	& 0.881 &	0.593 &	0 &	0.5 \\ \hline
Infiltration &	0.54 &	0.14 &	0.712 &	0.512 &	0.193 &	0.916 &	0.643 &	0.303 &	0.5 \\ \hline
Mass &	0.167 &	0.143 &	0.9	 & 0.044 &	0.136 &	0.727 &	0.529 &	0.236 &	0.5 \\ \hline
Nodule	& 0.595 &	0.129 &	0.352 &	0.631 &	0.123 &	0.868 &	0.484 &	0.183 &	0.5 \\ \hline
Atelectasis	&	0.41 &	0.143 &	0.65 &	0.369 &	0.147 &	0.864 &	0.54 &	0.239 &	0.5 \\ \hline
Pneumothorax &	0.869 &	0.131 &	0 &	1 &	NaN	& 0.869 &	0.537 &	0 &	0.5 \\ \hline
Pleural\_Thickening &	0.862 &	0.138 &	0 &	1 &	NaN	& 0.862 &	0.581 &	0 &	0.5 \\ \hline
Pneumonia &	0.514 &	0.119 &	0.54 &	0.511	& 0.13 &	0.892 &	0.536 &	0.209 &	0.5 \\ \hline
Fibrosis & 0.855 &	0.145 &	0 &	1 &	NaN &	0.855 &	0.477 &	0 &	0.5 \\ \hline
Edema & 0.624 &	0.119 &	0.54 &	0.635 &	0.167 &	0.911 &	0.601 &	0.255 &	0.5 \\ \hline
Consolidation &	0.65 &	0.126 &	0.377 &	0.689 &	0.149 &	0.885 &	0.559 &	0.214 &	0.5 \\ \hline

\end{tabular}}
  \label{Tab:1000}
\end{table*}

\begin{figure}
\centering
\subfloat[Under-fit model.]{%
\resizebox*{7cm}{!}{\includegraphics{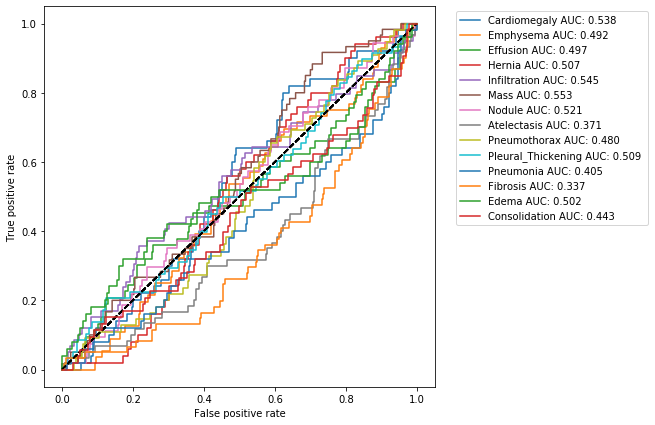}}}\hspace{5pt}
\subfloat[Final model.]{%
\resizebox*{7cm}{!}{\includegraphics{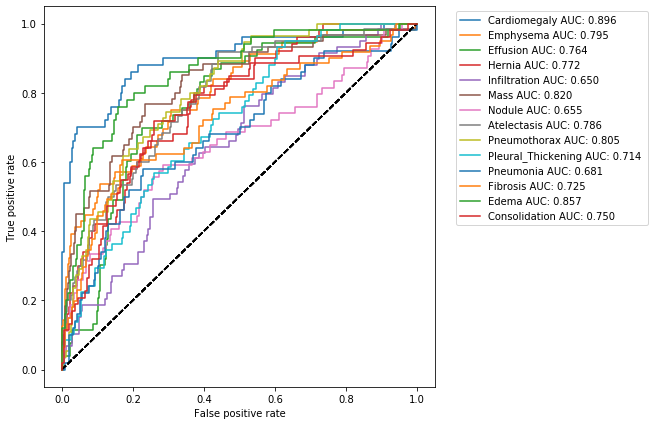}}}
\caption{ROC curve on classification of thoracic pathologies by a model trained on complete dataset.} \label{fig:ROCPlots}
\end{figure}

To address the limitations, we attempted two strategies. Firstly, we increased the size of the training dataset to 99,000 images while keeping the number of epochs constant. Unfortunately, this did not lead to an improvement in the generalization performance of the model. The ROC plot in Figure \ref{fig:ROCPlots}(a) shows the ROC curve below the diagonal, indicating that the model was still unable to capture the underlying patterns in the data. Our second strategy included increasing the number of training epochs to 300 and introducing regularization techniques for adaptive learning rate and dropout \cite{srivastava2014dropout} of 10\%. These techniques were implemented to prevent overfitting on the training data and improve the model's ability to capture more complex patterns.

The ROC curve shown in Figure \ref{fig:ROCPlots}(b) illustrates the improved performance of the classifier for different pathology conditions across various thresholds. The corresponding evaluation metric table in Table \ref{Tab:regualrize} provides the final measures of evaluation criteria for each condition. The results demonstrate a significant improvement in the discriminative performance of the model compared to our previous experiments. {Specifically, the substantial improvement in model performance when moving from 1,000 images to 99,000 images with large training epochs indicates that data quantity plays a critical role in training deep CNN models. The model also did not overfit the data during the extended training period of 300 epochs. }

In particular, the model achieved high sensitivity and accuracy on all conditions, indicating its ability to correctly identify positive cases. However, it is important to note that the positive predictive value (PPV) of the predictions can still be low. For example, for the Pneumonia condition, the sensitivity is 0.6, but given that the model predicted a positive result, the probability that the person actually has Pneumonia is only 0.18. These results highlight the importance of considering both sensitivity and PPV in evaluating the performance of the model.

\begin{table*} [h!]
    \caption{Table of evaluation metrics after training on 99000 images with regularizer}
    \resizebox{\textwidth}{!}{\begin{tabular}{l c c c c c c c c c}
    \hline
 & \textbf{Accuracy} &	\textbf{Prevalence} & \textbf{Sensitivity}	& \textbf{Specificity} &	\textbf{PPV}	 & \textbf{NPV} & \textbf{AUC} &	\textbf{F1} &	\textbf{Threshold}
\\ \hline

Cardiomegaly  & 0.826 &	0.119 &	0.78 &	0.832 &	0.386 &	0.966 &	0.896 &	0.517 &	0.5 \\ \hline
Emphysema  &	0.762 &	0.133 &	0.589 &	0.788 &	0.3	 & 0.926 &	0.795 &	0.398 &	0.5 \\ \hline
Effusion  &	0.681 &	0.126 &	0.736 &	0.673 &	0.245 &	0.946 &	0.764 &	0.368 &	0.5 \\ \hline
Hernia  &	0.705 &	0.119 &	0.66 &	0.711 &	0.236 &	0.939 &	0.772 &	0.347 &	0.5 \\ \hline
Infiltration &	0.598 &	0.14 &	0.644 &	0.59 &	0.204 &	0.91 &	0.65 &	0.31 &	0.5 \\ \hline
Mass  &	0.764 &	0.143 &	0.75 &	0.767 &	0.349 &	0.948 &	0.82 &	0.476 &	0.5 \\ \hline
Nodule  &	0.66 &	0.129 &	0.593 &	0.669 &	0.209 &	0.918 &	0.655 &	0.309 &	0.5 \\ \hline
Atelectasis	 &	0.674 &	0.143 &	0.75 &	0.661 &	0.269 &	0.941 &	0.786 &	0.396 &	0.5 \\ \hline
Pneumothorax &	0.7 &	0.131 &	0.745 &	0.693 &	0.268 &	0.948 &	0.805 &	0.394 &	0.5 \\ \hline
Pleural\_Thickening  &	0.6 &	0.138 &	0.672 &	0.588 &	0.207 &	0.918 &	0.714 &	0.317 &	0.5 \\ \hline
Pneumonia &	0.633 &	0.119 &	0.6	 & 0.638 &	0.183 &	0.922 &	0.681 &	0.28 &	0.5 \\ \hline
Fibrosis &	0.65 &	0.145 &	0.623 &	0.655 &	0.235 &	0.911 &	0.725 &	0.341 &	0.5 \\ \hline
Edema  &	0.783 &	0.119 &	0.8 &	0.781 &	0.331 &	0.967 &	0.857 &	0.468 &	0.5 \\ \hline
Consolidation  &	0.607 &	0.126 &	0.792 &	0.58 &	0.214 &	0.951 &	0.75 &	0.337 &	0.5 \\ \hline

\end{tabular}}
  \label{Tab:regualrize}
\end{table*} 

\begin{table}[]
\centering
\caption{Comparing performance of the proposed architecture against single-label prediction models. (-) indicates the unavailability of the model prediction on that pathology.}
\label{tab:my-table-single}
\resizebox{\textwidth}{!}{%
\begin{tabular}{@{}lccccc@{}}
\toprule
Pathology    &  \cite{crosby2020deep}  &  \cite{taylor2018automated} &  \cite{cha2019performance} &  \cite{islam2017abnormality} & Ours \\ \midrule
Cardiomegaly & -             & -             & -          & 0.87  & 0.90             \\
Nodule       & -             & -             & 0.73      & -     & 0.66             \\
Pneumothorax & 0.80         & 0.75          & -          & -     & 0.81             \\ \bottomrule
\end{tabular}%
}
\end{table}

{In Table \ref{tab:my-table-single} and Table \ref{tab:my-table-mult}, we compare the performance of our proposed model against single and multi-label prediction models for selected pathologies. Table \ref{tab:my-table-single} shows that our proposed multi-label approach was able to outperform single-label models. In Table \ref{tab:my-table-mult}, the results indicate that our proposed architecture outperforms  Wang et al. \cite{wang2017chestx} and Irvin et al. \cite{irvin2019chexpert} in multiple detection whereas betters performance of  CheXNext \cite{chexnext2018}, which is the state-of-the-art chest x-ray disease prediction model, for cardiomegaly condition only.}

\begin{table}[]
\centering
\caption{Comparing performance of the proposed architecture against multi-label prediction models. (-) indicates the unavailability of the model prediction on that pathology.}
\label{tab:my-table-mult}
\resizebox{0.95\textwidth}{!}{%
\begin{tabular}{@{}lcccc@{}}
\toprule
Pathology     & \cite{wang2017chestx} & \cite{irvin2019chexpert} &  \cite{chexnext2018} & Ours  \\ \midrule
Cardiomegaly  & 0.81       & 0.85        & 0.83    & 0.90         \\
Edema         &  -           & 0.93        & 0.92    & 0.86          \\
Mass          & 0.71       & -            & 0.91    & 0.82           \\
Consolidation & 0.71       & 0.94        & 0.89    & 0.75             \\
Pneumothorax  & 0.81       & -            & 0.94    & 0.81          \\ \bottomrule
\end{tabular}%
}
\end{table}

\subsection{Model interpretation}

Interpreting deep learning models is a challenging task due to their complex architecture. Class Activation Maps (CAMs) \cite{kwasniewska2017deep} have emerged as a popular method for generating visual explanations of model predictions, particularly for convolutional neural networks (CNNs). CAMs provide insights into where the model focuses when making a classification, aiding in model interpretation.

In our study, we utilized GRADCAM, a technique based on CAMs, to generate heatmaps for highlighting important regions in X-ray images during prediction. While GRADCAM does not provide detailed explanations for the model's reasoning, it can be valuable for expert validation, allowing experts to assess whether the model is attending to relevant regions in the image for making associated predictions.

\begin{figure}[h]
\centering
\includegraphics[width=0.75\textwidth]{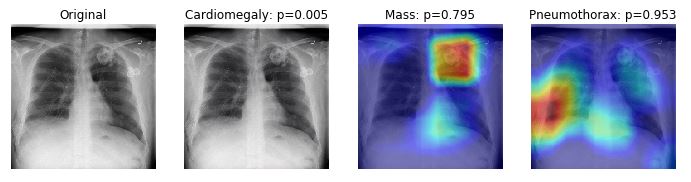}
\caption [Visualization of pathology prediction using GRADCAM] {Visualization of pathology prediction using GRADCAM}
\label{fig:heatmap}
\end{figure}

Figure \ref{fig:heatmap} presents the visualization of pathology prediction using GRADCAM. We generated heatmaps for two classes with the highest-performing AUC measures and one class with the lowest AUC measure. The activation map generates a heatmap that highlights the region of interest identified by the model when making predictions for Mass and Pneumothorax. The heatmap indicates the areas where the model is focusing its attention, providing insights into the features it considers important for the classification task. On the other hand, since the prediction for Cardiomegaly was very low, no heatmap is produced on the X-ray image, suggesting that the model did not identify any specific region as crucial for making a correct decision.


\subsection{Uncertainty estimation}

Computing the confidence interval of a measurement is a valuable technique that allows us to understand the impact of sampling and provides an estimation of uncertainty in our predictions based on the training dataset. It quantifies our confidence in the reliability of the model's predictions by considering that the training dataset is only a sample of real-world data.

In our experiment, we computed the confidence interval of the AUC score for each pathology and obtained the results, as presented in Table \ref{tab:CI}. These findings provide valuable insights into the reliability and consistency of our model predictions. The confidence interval represents a range of values within which we can be confident that the true AUC score falls. The narrower the interval, the more precise and confident we are in our predictions.

\begin{table}[h!]
\tbl{Table of confidence interval for AUC score.}
{\begin{tabular}{lc} \toprule
 Pathology & Mean AUC (CI 5\%-95\%) \\ \midrule
 Cardiomegaly &	0.89 (0.86-0.92) \\ \hline
Emphysema &	0.79 (0.76-0.82) \\ \hline
Effusion &	0.76 (0.73-0.80) \\ \hline
Hernia &	0.77 (0.73-0.80) \\ \hline
Infiltration &	0.65 (0.62-0.68) \\ \hline
Mass &	0.82 (0.79-0.85) \\ \hline
Nodule	& 0.65 (0.60-0.70) \\ \hline
Atelectasis	& 0.79 (0.76-0.81) \\ \hline
Pneumothorax &	0.81 (0.77-0.83) \\ \hline
Pleural\_Thickening &	0.71 (0.68-0.74) \\ \hline
Pneumonia &	0.68 (0.62-0.73) \\ \hline
Fibrosis &	0.72 (0.68-0.76) \\ \hline
Edema &	0.86 (0.82-0.88) \\ \hline
Consolidation &	0.75 (0.72-0.78) \\ \bottomrule
\end{tabular}}
\label{tab:CI}
\end{table}

Observing the confidence intervals, we can see that they are relatively narrow for almost all classes, indicating a high level of confidence in the results. This suggests that our proposed approach is robust and produces consistent results regardless of the specific sample of the training dataset. Narrow confidence intervals imply that the model's performance is consistent and reliable, reinforcing our confidence in the generalization ability of the model.

\section{Limitations}\label{sec:limitations}
Deep learning models should not be considered as a replacement for clinical diagnosis by medical professionals. These models should be used as complementary tools to aid medical professionals in making more accurate diagnoses. It is also crucial to validate the accuracy and reliability of these models on diverse and representative datasets. Interpretability and explainability of deep learning models in medical imaging tasks remain challenging research areas.

The deep learning model presented in this study has several limitations that should be acknowledged. These limitations include:
\begin{enumerate}
    \item \textit{Limited Image Types:} The model was trained using only frontal X-ray images. The unavailability of other types of images, such as lateral radiographs, restricts the model's exposure to different views and perspectives of the thoracic region. Incorporating multiple image types could enhance the model's performance and its ability to capture a comprehensive understanding of various pathologies.
    \item \textit{Lack of Medical History:} The model solely relies on the information extracted from the X-ray images and does not consider the patient's medical history. Medical history, including patient symptoms, previous diagnoses, and other relevant clinical information, plays a crucial role in disease diagnosis. The model could potentially improve its accuracy and diagnostic capabilities by incorporating such contextual information.

    \item \textit{Evaluation by Medical Professionals:} While the model demonstrates high sensitivity and specificity, it is essential to emphasize that medical professionals have not evaluated it. Comparing the model's performance to human performance is challenging, as it requires expert evaluation and validation. Therefore, further assessment by medical professionals is necessary to establish the model's clinical utility and compare its performance to that of human experts.

\end{enumerate}

\section{Conclusion and future works}\label{sec:conclusion}
In conclusion, this study demonstrates the potential of deep neural network models in predicting thoracic diseases using chest X-ray images. The evaluation of the model's performance indicates its effectiveness in detecting various pathologies. Additionally, using interpretable techniques such as Class Activation Maps (CAMs) provides insights into the model's decision-making process and aids in expert validation.

However, there are still several areas for improvement and further research. Incorporating different types of X-ray images, such as lateral radiographs, could enhance the model's performance and expand its applicability. Integration of patient medical history and contextual information may improve the accuracy and diagnostic capabilities of the model. Furthermore, collaboration with healthcare professionals is necessary to validate the model's predictions and assess its performance compared to human experts.

The development of an automated diagnosis system that combines the power of deep learning models with interpretability and expert collaboration holds great potential in improving access to reliable and efficient disease detection. By leveraging these technologies, we can work towards the goal of providing accurate and accessible diagnostic tools for thoracic diseases, ultimately benefiting patients worldwide and improving healthcare outcomes.

\subsection*{Data Availability}
The dataset is publicly available at NIHCC website\footnote{\url{https://nihcc.app.box.com/v/ChestXray-NIHCC}}.

\subsection*{Funding}
No funding was received to assist with this manuscript's research work or preparation.

\subsection*{Conflict of interest}
The authors have no competing interests to declare that are relevant to the content of this article.

\subsection*{Ethical approval}
This article does not contain any studies with human participants performed by any of the authors. The article uses open-source datasets.

\bibliographystyle{tfcse}
\bibliography{interactcsesample}
\end{document}